\documentclass{article}

\usepackage{PRIMEarxiv}

\usepackage[utf8]{inputenc} 
\usepackage[T1]{fontenc}    
\usepackage{hyperref}       
\usepackage{url}            
\usepackage{booktabs}       
\usepackage{amsfonts}       
\usepackage{nicefrac}       
\usepackage{microtype}      
\usepackage{lipsum}
\usepackage{fancyhdr}       
\usepackage{graphicx}       
\graphicspath{{media/}}     

\title{Resonance energy and wave functions of $^{31}$Ne: a calculation using supersymmetric quantum mechanics
\thanks{\textit{\underline{Citation}}: 
\textbf{M. Hasan, Md. A. Khan* (*Corresponding author). Title. Pages.... DOI:000000/11111.}} 
}

\author{M. Hasan (1) and Md. A. Khan* (2)\\
 (1,2)Department Of Physics \\
 Aliah University \\
Newtown, Kolkata, India\\
  \texttt{$^{*}$drakhan.rsm.phys@gmail.com, drakhan.phys@aliah.ac.in} \\
}

\begin{document}
\maketitle

\begin{abstract}
In this communication, we present an efficient method for computation of energy and wave function of weakly bound nuclei by the application of supersymmetric quantum mechanics (SSQM) and bound states in continuum (BIC) technique. As a case study the scheme is implemented to the two-body ($^{30}$Ne + n) cluster model calculation of neutron-rich nucleus $^{31}$Ne. Woods-Saxon central potential with spin-orbit component is used as the core-nucleon interaction. The two-body Schr\"{o}dinger equation in relative coordinate is solved numerically to get the energy and wave function of the low-lying bound states. A one-parameter family of isospectral potential (IP) is constructed from the bound state solutions following algebra of SSQM to find energies and wave functions of the resonance states. In addition to the 2p$_{3/2^-}$ (-0.33 MeV) ground state, two bound excited states: s$_{1/2}$ (-0.30 MeV), $p_{1/2}$ (-0.15 MeV) are also obtained. Few low-lying resonance states: f$_{7/2_1}$ (2.57 MeV), f$_{7/2_2}$ (4.59 MeV), f$_{5/2_1}$ (5.58 MeV), p$_{1/2_1}$(1.432 MeV), p$_{1/2_2}$ (4.165 MeV), p$_{3/2_1}$ (1.431 MeV), p$_{3/2_2}$ (4.205 MeV) are predicted. Among the predicted resonance states, the f$_{7/2^{-}}$ state having resonance energy $E_R \simeq 4.59$ MeV is in excellent agreement with the one found in the literature.
\end{abstract}


\section{Introduction}\label{1}
One of the most exciting areas of research in nuclear physics involves exotic nuclei that appear away from the nuclear stability line but closer to the neutron (n) or proton (p) driplines in the nuclear chart. Limiting lines on either side of the nuclear stability line, along which the single nucleon separation energy is zero, are called nuclear driplines. Panin {\it et al.} 2021 \cite{Panin-2021} have recently commented on the significance of the dripline nuclei in areas nuclear physics, nuclear medicine, nuclear astrophysics, defence research, material science, etc. The discovery of halo nuclei near the drip lines after the advent of radioactive ion beam (RIB) facilities is considered one of the most significant breakthrough events in nuclear physics. Typical halo nuclei consist of a dense core having a low-density envelope of the loosely bound nucleon(s). These weakly bound halo nuclei seldom have any excited bound state except the ground state having energy typically $<$ 1 MeV, near the continuum. Halo nuclei have matter radii more than the liquid drop model (LDM) prediction of $R_{A}\propto A^{1/3}$, small valence nucleon(s) separation energies, and high probability of occupation to low-$l$ shells. Another exciting and scientifically valued characteristic of halo nuclei is their resonance state(s) just above the binding threshold \cite{Kasuya-2021, Shubhchintak-2014, Liu-2002}.

Tanihata {\it et al.} 1985 \cite{Tanihata-1985} is credited for the first confirmation of 2n-halo structure in neutron-rich $^{11}$Li. One-neutron halo has been affirmed in $^{11}$Be\cite{Fukuda-1991, Schmitt-2012}, $^{19}$C\cite{Nakamura-1999, Yamaguchi-2011, Khan-2021}, $^{31}$Ne\cite{Nakamura-2014, Shi-2014, Watanabe-2014} while two-neutron halo are observed in $^{6}$He\cite{Tanihata-1992, Dutta-2003}, $^{11}$Li\cite{Tanihata-1985, Souza-2016}, $^{14}$Be\cite{Ibraheem-2019} and in $^{22}$C\cite{Togano-2016, Hasan-2019}. Nuclei $^{8}$B\cite{Fukuda-1999}, $^{26}$P\cite{Ren-1996}, $^{17}$F\cite{Ibraheem-2019}, etc have one proton halo structure while $^{27}$S \cite{Cai-2002}, $^{17}$Ne\cite{Tanaka-2005}, etc have two-proton halo structure. The neon isotope $^{31}$Ne has been under scanner due to its position in the mixing zone of normal and intruder shell configurations. Audi {\it et al.} 2003 \cite{Audi-2003} hinted a possible halo structure of $^{31}$Ne with one neutron separation energy $S_{n} \simeq 0.33$ MeV. Studies of Urata {\it et al.} 2011 \cite{Urata-2011}, Takechi {\it et al.} 2012 \cite{Takechi-2012} also indicated a p- or a s-state halo structure in $^{31}$Ne, consistent with the findings of a Coulomb-breakup experiment at RIKEN's Radio-Active Ion Beam Factory (RIBF) by Nakamura {\it et al.} 2009 \cite{Nakamura-2009}. They measured the one-neutron removal cross section in $^{31}$Ne and suggested halo structure with spin-parity J$^{\pi} = 3/2^-$ contrary to the shell model prediction of $7/2^-$. On the other hand, the configuration $^{30}$Ne $(0_{1}^{+})\otimes$2p$_{3/2}$ of spin-parity $3/2^-$ is compatible with the energy data S$_{n}\sim 0.4$ MeV. The $^{30}$Ne $(0_{1}^{+}) \otimes$ 2s$_{1/2}$ with J$^{\pi} = 1/2^+$ is also compatible with the data for S$_{n} \simeq 0.8$ MeV. The observed large Coulomb breakup cross section of $540 (70)$ mb for $^{31}$Ne is indicative of a soft E1 excitation and $1/2^+$ or $3/2^-$ spin-parity of the ground state, though majority of them found $3/2^-$ as the spin-parity of the ground state. Jurado {\it et al.} 2007\cite{Jurado-2007} first noticed halo structure in $^{31}$Ne through direct mass measurements and obtained S$_{n} = 0.29\pm 1.64$ MeV. Wapstra {\it et al.} 2003 \cite{Wapstra-2003} estimated S$_n = 0.332 \pm 1.069$ MeV. Gaudefroy {\it et al.} 2012\cite{Gaudefroy-2012} confirmed one neutron halo in $^{31}$Ne with valence neutron occupying the 2p$_{3/2}$ orbital. 

According to shell-model scheme, $^{31}$Ne ground state should have $^{30}$Ne$(0^+)\otimes$ 1f$_{7/2}$ configuration. But, Poves and Retamosa 1994 \cite{Poves-1994}, Descouvemont 1999 \cite{Descouvemont-1999} predicted a possible inversion between the shell model levels 1f$_{7/2}$ and 2p$_{3/2}$ for the ground state of $^{31}$Ne. They predicted valance neutron occupying the 2p$_{3/2}$ intruder orbit with a dominant $^{30}Ne (0^{+})\otimes$ p$_{3/2}$ single-particle configuration. Ren {\it et al.} 2001\cite{Ren-2001} predicted $3/2^-$ as the ground state of $^{31}$Ne instead of shell model label $7/2^{-}$ using DDRMF concept. Hamamoto 2010\cite{Hamamoto-2010} using scattering phase shift method found 1f$_{7/2}$ resonance of width 0.224 MeV at 2.40 MeV. Using the particle-rotor model, Urata {\it et al.} 2011 \cite{Urata-2011} found the ground state configuration of $^{31}$Ne as J$^{\pi}=3/2^-$. Liu {\it et al.} 2012\cite{Liu-2012} reported the energy and widths of low-lying neutron resonances in $^{31}$Ne using a complex scaling technique and identified three resonances namely 1f$_{7/2}$, 1f$_{5/2}$, and 1g$_{9/2}$ among which the acceptable one is the $\frac{7}{2}^-$ resonance close to $4.592$ MeV. Using analytic continuation approach for resonances Zhang {\it et al.} 2014\cite{Zhang-2014} predicted one-neutron p-orbit halo structure in $^{31}$Ne. By complex momentum representation technique, Tian {\it et al.} 2017\cite{Tian-2017} found $2p_{3/2}$ resonance together with 1f$_{7/2}$ resonance, as well as a $p-f$ inversion in single-particle levels. Very recently, He 2019\cite{He-2019} using halo effective field theory (HEFT) with effective range method obtained S$_n$ for the ${3/2}^-$ ground state and predicted ${1/2}^+$ state in-line with the findings of Nakamura {\it et al.} 2014\cite{Nakamura-2014}.

Literature survey indicated a $^{30}$Ne($0^+)\otimes$p$_{3/2}$ ground configuration of $^{31}$Ne instead of the shell model configuration $^{30}$Ne($0^+)\otimes f_{7/2}$. However, in this work all possible configurations for the bound and resonance states of $^{31}$Ne will be explored in the framework of ($^{30}$Ne +n) two-body cluster model. Resonance energy computation in weakly bound systems can be accomplished by the application of supersymmetric quantum mechanics (SSQM). The scheme is based on the fact that for any arbitrarily chosen potential (say, {$V$}), one can generate a one-parameter family of isospectral potential (IP) ($\hat{V}$), with a free parameter ($\eta$). The present method has an edge over several other methods providing resonance properties, namely- the Complex Scaling Method reported by Moiseyev 1998 \cite{Moiseyev-1998}, the R-matrix method of Descouvemont and Vincke 1990 \cite{Descouvemont-1990}, the ACCC method of Kukulin {\it et al.} 1989 \cite{Kukulin-1989}, the Numerov method reported by Baluja {\it et al.} 1982\cite{Baluja-1982} or the one using phase shifts to extract the resonance parameters, because of the advantages of the robustness of the algebra of supersymmetric quantum  mechanics. When the original potential has a shallow well following a wide skinny barrier (poorly supporting any resonance state), $\eta$ can be selected judiciously to increase the depth of the well and height of the barrier (in $\hat{V}$) simultaneously. It is accomplished by decreasing the value of $\eta$ in suitable steps and checking its effect to the potential. The optimized well-barrier combination in $\hat{V}$ facilitates trapping of the particle inside it, resulting in an accurate computation of the resonant state exactly at the same energy, as that in the original shallow potential {$V$}. This is because, {$V$} and $\hat{V}$ are {\bf strictly isospectral}. Earlier, the scheme has been implemented successfully in two-body model calculation for $^{19}$C by Khan {\it et al.}, 2021\cite{Khan-2021}, in three-body cluster model calculation of resonances in $^{22}$C by Hasan {\it et al.}, 2019\cite{Hasan-2019} and in $^{42,44}$Mg by Khan {\it et al.} 2021\cite{Khan2-2021}. 

\section{Theoretical Scheme}\label{2}
For core-nucleon two-body model of $^{A_c+1}$X ($\equiv ^{A_c}$X+N) nucleus, the relative motion can be described by the Schr\"{o}dinger equation
\begin{equation}\label{eq1}
\left[-\frac{\hbar^{2}}{2\mu}\frac{d^{2}}{d r^{2}}
+\frac{L(L+1)\hbar^2}{2\mu r^2} + V(r)-E\right]\psi(r)=0.
\end{equation}
where $\mu=\frac{(A_cm)(m)}{A_cm+m}=\frac{A_cm}{A_c+1}$ = reduced mass of the core-N two-body system,  $m$ is the nucleon mass and $V(r)$ is the core-nucleon potential. In terms of the effective potential, 
\begin{equation}\label{eq2}
U(r)= V(r)+\frac{\hbar^2}{2\mu}\frac{L(L+1)}{r^2},
\end{equation}
 Eq.\,(1) can be rewritten as 
\begin{equation}\label{eq3}
\left\{-\frac{\hbar^{2}}{2\mu}\frac{d^{2}}{dr^{2}} + U(r) - E \right\} \psi(r) = 0. \end{equation}\label{eq2}
Numerical solution of Eq.\,(3) by the re-normalized Numerov method (RNM) prescribed by Johnson 1978 \cite{Johnson-1978} subject to appropriate boundary conditions yield the ground state energy $E$ (=$E_0<0$) and wave function $\psi_0(r)$.

\subsection{Construction of one-parameter family of IP using SSQM}\label{2.1}
After obtaining the ground state energy ($E_0$) and wave function ($\psi_0(r)$), a one-parameter family of IP \begin{equation}\label{eq4}
\hat{U}(\eta; r)= U(r)-E_0 - 2\frac{\hbar^{4}}{2\mu}
\frac{d^{2}}{dr^{2}}\log [I_0(r) + \eta]
\end{equation}
can be constructed following Nieto 1984\cite{Nieto-1984}, Khare and Sukhatme 1989\cite{Khare-1989}, and  Cooper {\it et al.} 1995 \cite{Cooper-1995} applying formalism of SSQM. In Eq.\,(4), $\eta$ is a free parameter and 
$I_0(r)={\displaystyle\int}_{r^{\prime}=0}^{r} {[\psi_{0}(r^{\prime})]}^{2} dr^{\prime}$; $\int_0^{\infty}|\psi_0(r)|^2dr =1$. The parameter $\eta$ is arbitrary in the intervals $-\infty<\eta<-1$ and $0<\eta<\infty$. As $I_{0}(r)$ lies between 0 and 1, the interval $-1\leq \eta\leq 0 $ is forbidden, to bypass singularities in $\hat{U}(\eta; r)$. For $\eta \rightarrow \pm \infty$, $\hat{U} \rightarrow U$ and for $\eta \rightarrow 0+$, $\hat{U}$ develops a narrow and deep attractive well near the origin following a high barrier. This improved well-barrier combination facilitates trapping of the particle giving rise to sharp resonance at the same energy as that in the original effective potential $U(r)$. However, a delta-function like behaviour of the improved well-barrier combination may add large computation error in the observables, hence a judicious choice of $\eta$ is important. The above procedure holds for resonance state of same spin-parity ($J^{\pi}$) as that of the bound state but fails for states of different $J^{\pi}$. In the latter case one needs to combine the bound states in continuum (BIC) technique of Pappademos {\it et al.} 1993\cite{Pappademos-1993} with the formalism of SSQM to accomplish the desired goal. \\
When the original effective potential of Eq.\,(2) has a shallow well followed by a low and wide barrier, an accurate computation of energy and width is not feasible numerically. Though, in principle, for such a barrier of finite height, a system may be temporarily trapped inside the well, when its energy matches the resonance energy. But, in these cases penetrability through the barrier will be large enough to give a broad resonance width. In our scheme, such numerical obstacles can be bypassed and a reliable energy and width can be obtained by generating a one parameter family of IP having optimum depth to suppress the tunnelling probability of the system, yet providing the correct position of the resonance state. To accomplish this, Eq.\,(3) is solved for some positive energies E ($>0$) subject to the initial condition $\psi(0) = 0$. This positive energy $\psi(r)$ is non-normalizable and oscillates with constant amplitude in the asymptotic region where the potential $U(r)$ vanishes. Using this $\psi(r)$ one can readily construct the potential 
\begin{equation}\label{eq5}
\hat{U}(\eta;r) = U(r) - \frac{\hbar^2}{2\mu}\left[\frac{4\psi(r)\psi^{\prime}(r)}{I(r)+\eta} - \frac{2\psi(r)^4}{(I(r)+\eta)^2}\right]
\end{equation}
where $I(r)={\displaystyle\int}_{r^{\prime}=0}^{r} {[\psi(r^{\prime})]}^{2} dr^{\prime}$. Eq.\,(3) has the solution $\hat{\psi}(\eta; r)= \frac{\psi(r)}{I(r)+\eta}$ at the same energy $E$ when $\hat{U}(\eta; r)$ replaces $U(r)$. As $\psi(r)$ is non-normalizable and oscillates in the asymptotic region, $I(r)$ grows approximately linearly with $r$ making $\hat{\psi}(\eta; r)$ normalizable at larger $r$ as described by Pappademos {\it et al.}\cite{Pappademos-1993}. In this way, $\hat{\psi}(\eta; r)$ represents a bound state in  continuum (BIC) of $\hat{U}(\eta; r)$, since $\hat{U}(\eta; r)$ is strictly isospectral with $U(r)$. BIC is a less-familiar result of SSQM\cite{Darboux-1882, Nieto-1984, Khare-1989, Pappademos-1993}. For $\eta \rightarrow \pm \infty$, $\hat{U} \rightarrow U$ and for $\eta \rightarrow 0+$, $\hat{U}$ develops a narrow and deep attractive well near the origin following a high surface barrier as can be seen from the data presented in Table 2 (see later) and also from the schematics of Fig.\,3 (see later). As the improved well-barrier combination effectively traps the system resulting in a quasi-bound state, we may define a trapping probability as \begin{equation}\label{eq6}
\xi(E)=\int_{r^{\prime}=0}^{r_B} |\hat{\psi}(\eta; r^{\prime})|^2 dr^{\prime}
\end{equation}
where $r_B$ indicates the location of the top of the potential barrier. A plot of $\xi(E)$ versus $E$ ($E > 0$) shows prominent peak at the resonance energy, $E = E_R$. WKB approximation is used to compute resonance width ($\Gamma$) in terms of the barrier transmission coefficient and time of flight between the classical turning points corresponding to $E_R$ in the potential-well represented by $\hat{U}(\eta; r)$ (see Eq.\,(5)):
\begin{equation}\label{eq7}
\Gamma = 2\sqrt{\frac{\hbar^2}{2\mu}}  \frac{\exp [-2\int_b^c\sqrt{\frac{2\mu}{\hbar^2}\left\{\hat{U}(\eta; r) - E_R\right\}} dr]}{\int_a^b \frac{dr}{\sqrt{\left\{E_R-\hat{U}(\eta; r)\right\}}}}
\end{equation} 
In Eq.\,(7), $a, b, c$ represent the classical turning points corresponding to $E_R$ in the potential of Eq.\,(5) for the chosen $\eta$ satisfying the condition: $\hat{U}(\eta; a)=\hat{U}(\eta; b)=\hat{U}(\eta; c) = E_R$.
\section{Application to $^{31}$Ne}\label{3}
The above scheme is tested against bound and resonance states of $^{31}$Ne which has weakly bound excited state of spin-parity J$^{\pi} = 1/2^+$ in addition to the $3/2^-$ ground state (see Nakamura {\it et al} 2014 \cite{Nakamura-2014} and He 2019 \cite{He-2019}). The system also exhibits low-lying resonances of spin-parity J$^{\pi} = 7/2^-$ as reported by Hamamoto 2010\cite{Hamamoto-2010} and Liu {\it et al.} 2012\cite{Liu-2012}. The nucleus $^{31}$Ne being strong one-neutron halo candidate can be treated as a two-body system consisting of a structure-less core (point like particle) plus one valence neutron moving around the $^{30}$Ne core. Hence, $3/2^-$ ground state may be regarded as a p-wave state, while the predicted $1/2^+$ state may be regarded an s-wave state. These s- and p-wave states are the outcome of the interaction between valence neutron and core \cite{He-2019} nucleus. Here, Pauli restrictions arising due valence neutron and the core nucleons is suppressed by considering the core as point like particle. Chosen Woods-Saxon potential having a spin-orbit term, adopted from works of Pahlavani \cite{Pahlavani-2012} for the core-n pair is given by $V_{core-n}(r)= -V_{c}f(r) + V_{LS}\frac{1}{r}\frac{df(r)}{dr}(\bf L.S)$, where $f(r)=[1+\exp(\frac{r-R_c}{a})]^{-1}$; $V_{LS}=0.44V_c(R_0/\hbar)^2$ MeV; surface diffuseness parameter, $a = 0.60$ fm; $R = R_0 A^{1/3}$; $R_c$($^{30}$Ne) = $1.2\times 30^{1/3}$ fm = 3.73 fm. Here, we have three adjustable parameters \textit{viz.} $a, V_c$ and $R_0$. For $L = 0$, $U(r)$ develops an attractive well giving rise to the $J^{\pi}=1/2^+$ neutron bound state in $^{31}$Ne. 
For $L > 0$, the centrifugal term $\frac{31\hbar^2}{60m}\frac{L(+1)}{r^2}$ in $U(r)$ dominates for small $r$ and $U(r)$ exhibits an attractive well following a repulsive barrier as shown in Fig.\,1. This well-barrier combination effectively traps the system giving rise to resonance state(s) in addition to the bound state(s). Minimum of the core-n effective potential $U(r)$ for the bound $3/2^-$, $1/2^+$ and $1/2^-$ states are respectively -7.9 MeV, -5.4 MeV and -8.6 MeV (see Table 1). Energy and wave function of these states are obtained by numerical solution of Eq.\,(3) using RNM algorithm prescribed by Johnson 1978\cite{Johnson-1978}. Calculated bound state energies and root mean squared (RMS) matter radii have been listed in Table 1. 

For a fruitful demonstration of the resonances, we adopted the technique of the bound states in the continuum (BIC) introduced in the preceding section. BIC represents the solution of Eq.\,(3) for the IP $\hat{U}(\eta;r)$ represented by Eq.\,(5), in which $\eta$ manages the strength of $\hat{U}(\eta;r)$. It is seen that resonance energy is independent of $\eta$ and an appropriate choice of $\eta$, apart from protecting the stability of the resonant state, also preserves the spectrum of the original potential U(r), while adding a discrete BIC at specified energy. Here, we have constructed the IP following Eq.\,(5) for some of the low-lying unbound states: $1/2^-$, $3/2^-$, $5/2^-$, $7/2^-$ of $^{31}$Ne and obtained the corresponding resonant wave functions. The tuning parameter $\eta$ produced a {\bf dramatic effect} in the IP as indicated by the representative schematics in Fig.\,3 and data of Table 2 both obtained for the $7/2^-$ resonance state of $^{31}$Ne. Adjusted depth parameter $V_c$ together with the minimum of the original effective potential $U(r)$ for the resonance states are listed in columns 2 \& 3 of Table 3. Using the positive energy solution $\hat{\psi}(\eta; r)$, we have computed the probability of trapping $\xi(E)$, resonance energy ($E_R$) and resonance width ($\Gamma$). Computed resonance energies and widths are listed in columns 4 \& 5 of Table 3. A representative plot of the resonance state wave function is shown in Fig.\,4 for some arbitrary values of $\eta$. Schematic view of the resonance profiles for some arbitrary values of $\eta$ is shown in Fig.\,5. Similar nature of the resonance profiles are also seen for $1/2_{1}^-$, $1/2_{2}^-$, $3/2_{1}^-$ and $3/2_{2}^-$ resonances. Sharpness of resonances is found to increase with decreasing $\eta$ values. 

\section{Results and Discussions}\label{4}
Adjusted values of $V_c$ used to reproduce observed neutron separation energy S$_n$ for the low-lying states- $3/2^-$, $1/2^+$, $1/2^-$ of $^{31}$Ne are listed row-wise in Table 1 along with corresponding minimum of $U(r)$. Calculated $S_n$ together with some of them found in the literature are presented in Table 1. It is to be noted that $S_n$ for the state $1/2^-$ is a mere prediction only. The RMS matter radius calculated using wave functions of $3/2^-$, $1/2^+$, $1/2^-$ states are also listed Table 1. The original shallow potential ($\eta\rightarrow\infty$) represents a sufficiently broad and spatially extended resonance profile (not shown in the plot) as compared to the uppermost curve in Fig.\,5. Data in Table 2 reflects that smaller $\eta$ causes enhancement in the depth of the well and height of the barrier simultaneously. Position of the minimum of the well and maximum of the barrier are found to shift towards the origin as $\eta$ decreases. For sufficiently large $\eta$ ($\eta\rightarrow\infty$), the potential well has a depth of -6.79 MeV near 3.32 fm while the barrier has a height of 6.15 MeV near 3.73 fm. Which, for $\eta=5\times 10^{-5}$, changes to -268.81 MeV near 1.15 fm and 130.44 MeV near 1.79 fm respectively. These changes in the IP may be seen as a {\bf dramatic effect} in it (See Table 2). One of the major advantages of the present scheme is that the resonance energy ($E_R$) and width ($\Gamma$) are independent of $\eta$, except the computational error that creeps in $\Gamma$ as $\eta\rightarrow 0+$, due to delta-function like behaviour of the IP. Calculated energies and widths of the resonances are presented in columns 4 and 5 of Table 3, together with some of them found in the literature. Calculated energies of the bound states: $3/2^-$ (0.33 MeV), $\frac{1}{2}^+$ (0.30 MeV) presented in Table 1 are in excellent agreement with the experimental values within error bars. Calculated RMS matter radius of the bound state: $3/2^-$ (3.6 fm) also agrees fairly with the experimental values. Among the predicted low-lying resonances in $^{31}$Ne viz: f${_{7/2}}_1$ (E$_R$=2.570 MeV, $\Gamma=0.863\pm 0.087$), f${_{7/2}}_2$ (E$_R$=4.59 MeV, $\Gamma=0.803\pm 0.083$), p${_{1/2}}_1$(E$_R$=1.432 MeV, $\Gamma=0.866\pm 0.282$), p${_{1/2}}_2$ (E$_R$=4.165 MeV, $\Gamma=2.080\pm 0.565$), p${_{3/2}}_1$ (E$_R$=1.431 MeV, $\Gamma=0.828\pm 0.265$), p${_{3/2}}_2$ (E$_R$=4.205 MeV, $\Gamma=1.586\pm 0.523$), f${_{5/2}}_1$ (E$_R$=5.580 MeV, $\Gamma=0.821\pm 0.138$), the state f$_{7/2}$ having resonance energy E$_R$=4.59 MeV are in excellent agreement with the one found in the literature \cite{Liu-2012}.

\section{Conclusion}\label{5}
In this communication, we have reproduced the energy and wave functions of the low-lying bound and unbound states of $^{31}$Ne neutron halo nucleus using a spherically-symmetric two-body potential having a spin-orbit component. The SSQM formalism together with the BIC technique is successfully used to generate resonant state wave functions, resonance energy, and width of the resonance. The technique confirmed the existence of the unbound $7/2^-$ state in $^{31}$Ne and its resonance energy E$_R$ = 4.592 MeV\cite{Liu-2012}. In this technique, the parameter $\eta$ facilitates a fruitful demonstration of the resonance effect without affecting the exact location of the resonance energy or the position of the peak in the trapping probability versus energy plot as shown in Fig.\, 5. Physics of exotic nuclei formed near the drip-lines having quasi-bound or unbound states will continue to dominate the field of nuclear physics in the coming years. A sound theoretical framework involving less numerical uncertainties is inevitable for a meaningful description of their structure. Because, they are characterized by numerous types of long-lived (quasi-stationary) states, some of which include one particle shape resonances, one-particle virtual states, Effimov states, three-particle near-threshold long-lived states appearing due to the existence of bound, virtual or resonance states in their two-body subsystems, and compound states or quasi-bound states embedded in a continuum-an example of which is the Feshbach resonances in atomic physics. The SSQM scheme adopted in this work has an edge over others because, this is the only theoretical approach by which resonant state wave functions can be extracted and exploited to reproduce the experimental observable $\Gamma$. In our earlier works on exotic three-body systems-  $^{22}$C\cite{Hasan-2019}, $^{42,44}$Mg\cite{Khan2-2021}, we have applied this procedure successfully. The scheme has also been used for the unbound nucleus $^{15}$Be by Dutta {\it et al.} 2018\cite{Dutta-2018}. Thus, we may conclude that the present scheme applies also on two-body bound or unbound nucleus with excellent outcomes. 
 
\begin{figure}
\centering
\includegraphics[scale=0.75]{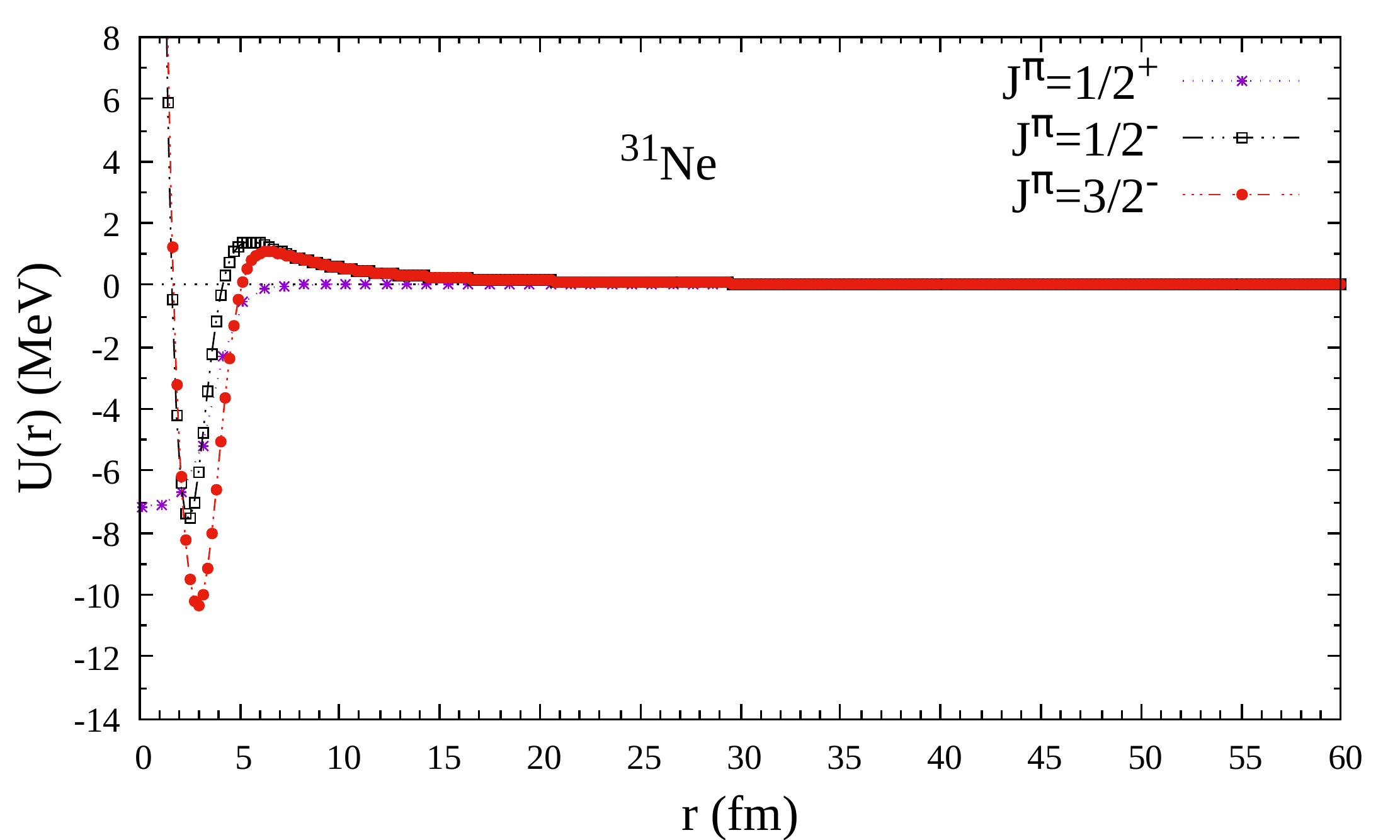}
\caption{Plot of original potential $U(r)$ as a function of radial distance $r$ for $1/2^{+}$, $1/2^{-}$ and $3/2^{-}$  bound states of $^{31}$Ne.}
\label{Fig:1}
\end{figure}
\begin{figure}
\centering
\includegraphics[scale=0.75]{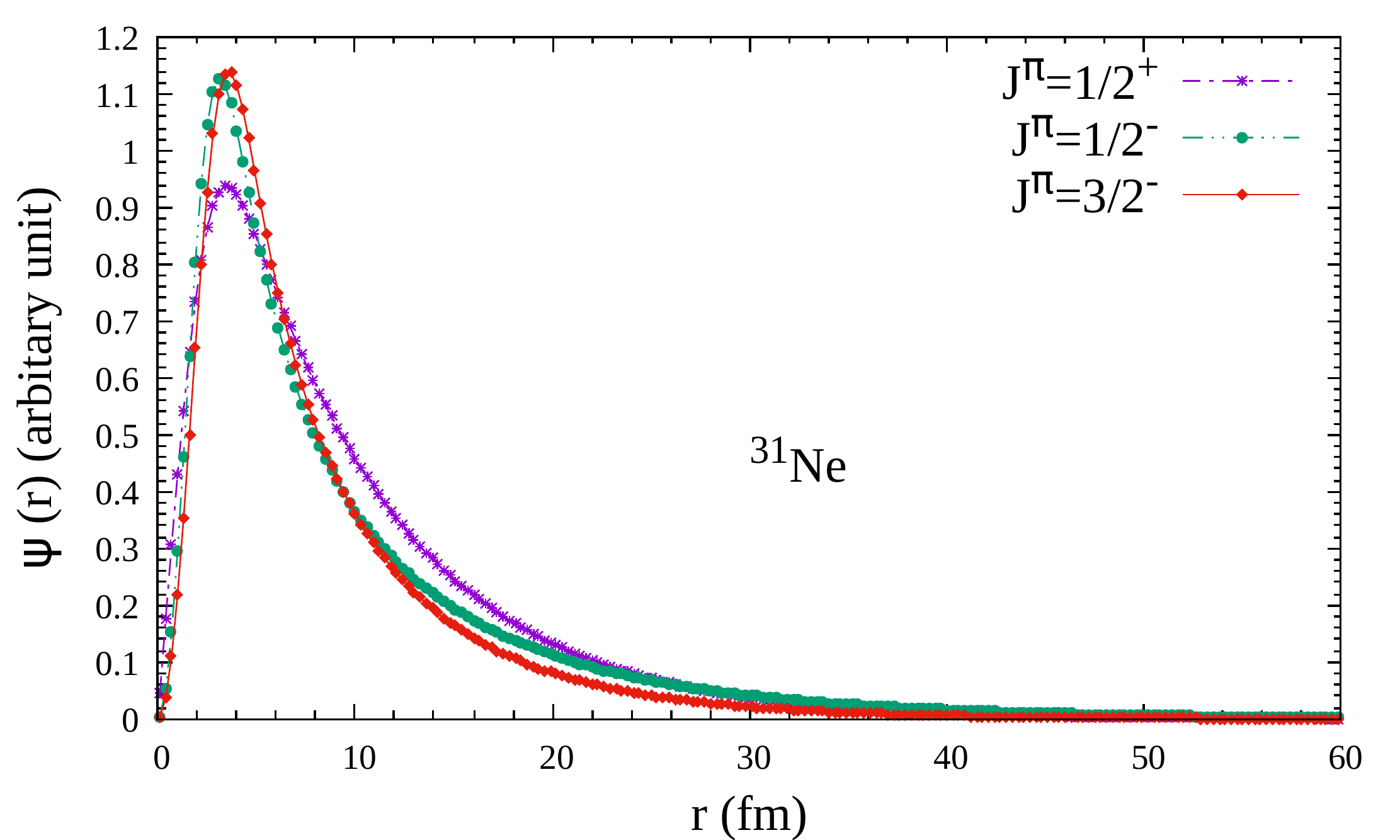}
\caption{Plot of wave function $\psi(r)$ as a function of radial distance $r$ for $1/2^{+}$, $1/2^{-}$ and $3/2^{-}$ bound states of $^{31}$Ne.}
\label{Fig: 2}
\end{figure}
\begin{figure}
\centering
\includegraphics[scale=0.75]{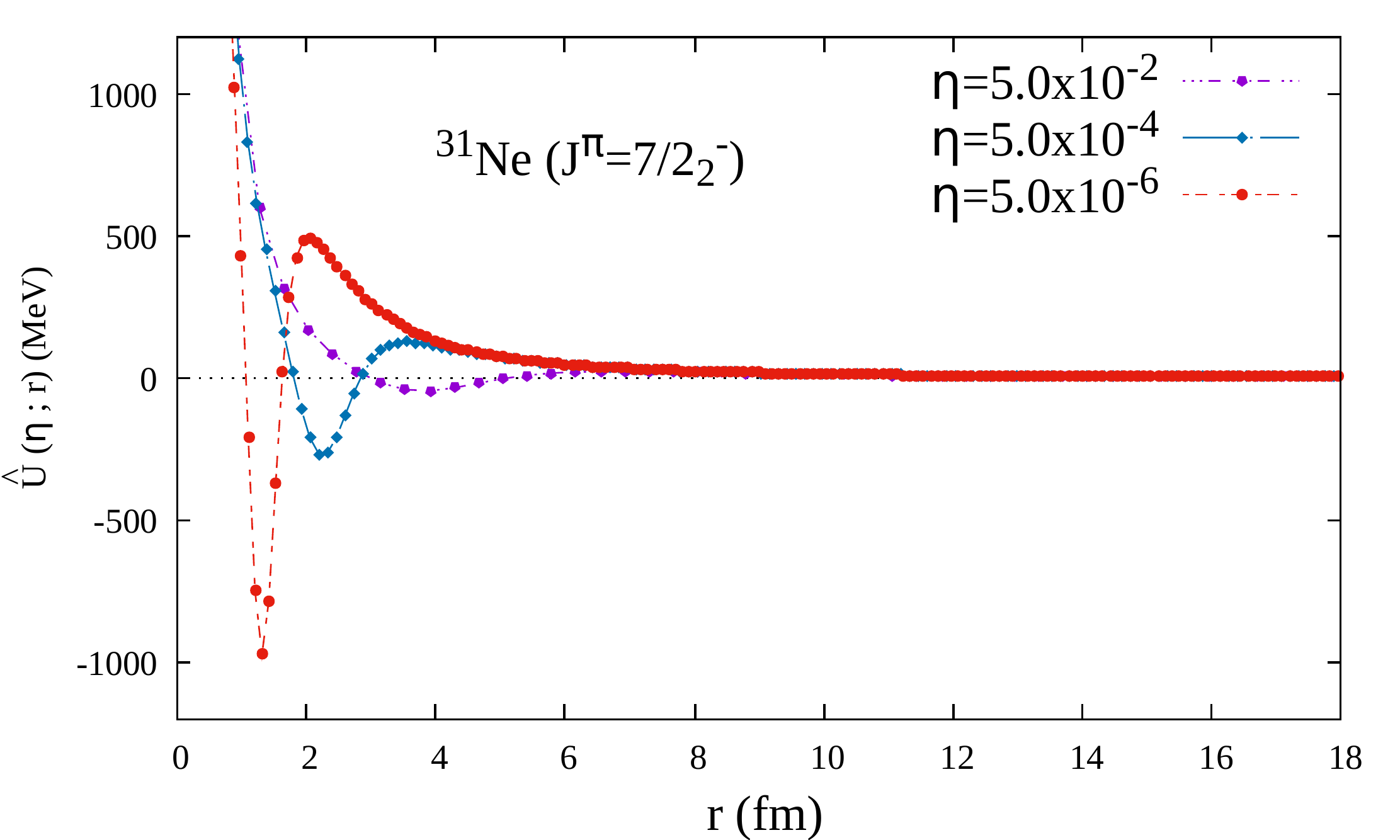}
\caption{Plot of isospectral potential (IP) $\hat{U}(\eta; r)$ as a function of radial distance $r$ for $7/2^{-}$ resonant state of $^{31}$Ne.}
\label{Fig: 3} 
\end{figure}
\begin{figure}
\centering
\includegraphics[scale=0.75]{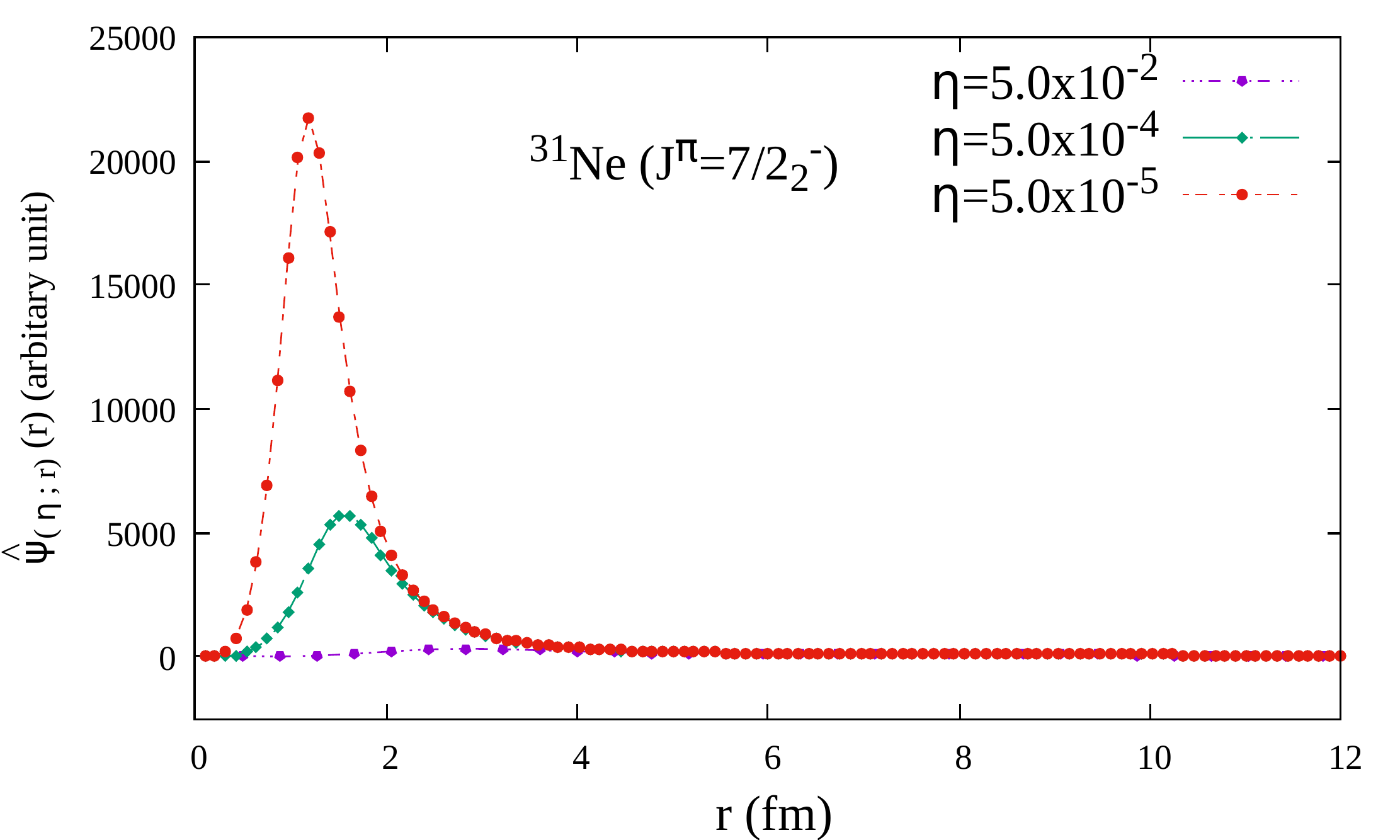}
\caption{Plot of resonant wave function ($\hat{\psi}(\eta; r)$) as a function of radial distance $r$ for $7/2^{-}$ resonant state of $^{31}$Ne corresponding to the resonance energy $E_{R} = 4.592$ MeV.}
\label{Fig: 4}
\end{figure}
\begin{figure}
\centering
\includegraphics[scale=0.75]{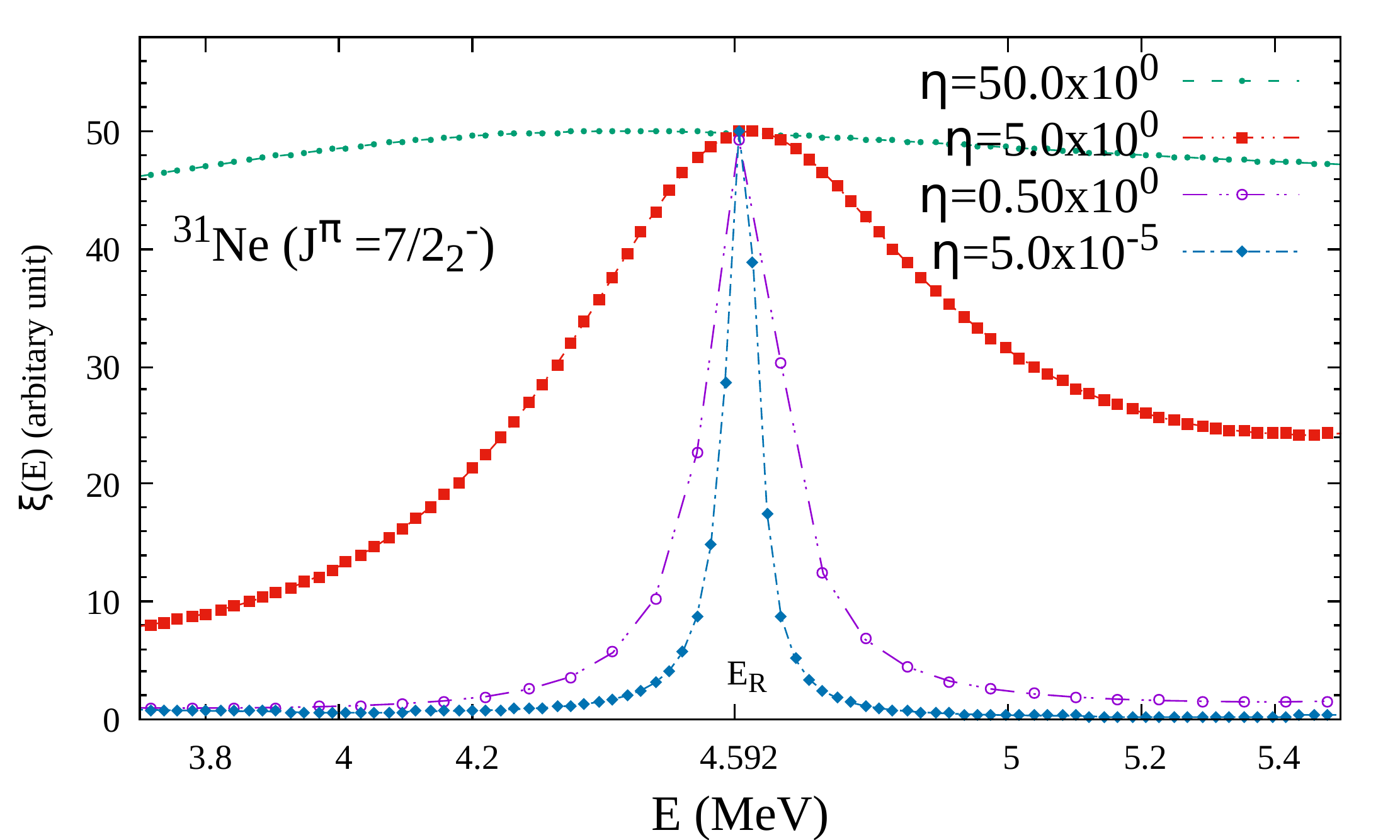}
\caption{Plot of the trapping probability $\xi(E)$ against energy E for $7/2^{-}$ resonant state of $^{31}$Ne.}
\label{Fig: 5}
\end{figure}
\begin{table}
\caption{Potential parameters, energies and RMS matter radius obtained for the low-lying states of $^{31}$Ne together with some of those found in the literature.}
\centering
\begin{tabular}{llll}\hline
\toprule
State,  &Gr. st.:  & Ex. st.: & Ex. st.: \\
$J^{\pi}$&$3/2^{-}$&$1/2^{+}$&$1/2^{-}$\\
\midrule
{\bf E (MeV)}&&&\\
$V_c$ &-15.1170&-5.4100&-18.9910\\
$U^{min}$&-7.9449&-5.3973&-8.6139\\
$S_{n}(Calc.)$&0.3317&0.3005 &0.1503\\
$S_{n}(Expt.)$&$0.15_{-0.10}^{+0.16}$\cite{Nakamura-2014}&$0.30_{- 0.17}^{+0.26}$\cite{Nakamura-2014}     &-      \\
&$0.332\pm1.069$\cite{Wapstra-2003}&&\\
{\bf Radius (fm)} &&&\\
R($^{31}$Ne)$_{Calc.}$ &3.6109&3.6090 &3.6094  \\
R($^{31}$Ne)$_{Expt.}$ & 3.47(AMD),\cite{Minomo-2012},     &&  \\
&3.62(AMD-RGM),\cite{Minomo-2012}&&\\
\bottomrule
\end{tabular}
\label{tbl1}
\end{table}

\begin{table}
\caption{Parameters of the isospectral potential (IP) for some representative values of $\eta$ derived for obtaining energy (E$_R$) of $7/2^-$ resonance state in $^{31}$Ne.}
\centering
\begin{tabular}{lllll}
\toprule
&\multicolumn{2}{l}{Potential Well}	&\multicolumn{2}{l}{Potential	Barrier}\\
\toprule
Tuning   &Depth&Pos. of minimum&Height&Pos. of maximum\\
factor,&V$_{w}$&r$_w$&V$_b$&r$_b$\\
$\eta$ &$(MeV)$&$(fm)$&$(MeV)$&$(fm)$\\
\midrule 
$\infty$ &-6.79 &3.32&6.15&5.73\\
50.00 &-9.01 &2.30&6.81 &5.39\\
$1\times 10^{-1}$  &-18.14 &2.70&8.86 &4.76\\
$5\times 10^{-2}$  &-24.03  &2.54&10.96 &4.48\\
$1\times 10^{-3}$  &-110.52&1.63&48.94&2.60\\
$1\times 10^{-4}$  &-221.97 &1.24&105.99&1.94\\
$5\times 10^{-5}$  &-268.81 &1.15&130.44&1.79\\
$5\times 10^{-6}$  &-489.03 &0.88&246.34&1.36 \\
\bottomrule
\hline
\end{tabular}
\label{tbl2}
\end{table}

\begin{table}
\caption{Comparison of the calculated results with those found in the literature for some low-lying resonant states in $^{31}$Ne.}
\centering
\begin{tabular}{lllll}\hline
\toprule
 Reso. & $V_{c}$&  $U_{min}$& $E_{r}$&Reso. width, $\Gamma$\\
 State $J^{\pi}$&  (MeV)& (MeV)& (MeV)& (MeV)\\
\noalign{\smallskip}\noalign{\smallskip}
\midrule
$1/2^{-}_1$&-18.99&-8.61&1.432&$0.866\pm0.282$\\
$1/2^{-}_{2}$&-18.99&-8.61&4.165&$2.080\pm0.565$\\
$3/2^{-}_{1}$&-15.12&-7.94&1.431&$0.828\pm0.265$\\
$3/2^{-}_{2}$&-15.12&-7.94&4.205&$1.586\pm0.523$\\
$5/2^{-}_{1}$&-51.31&-6.97&5.580&$0.821\pm0.138$\\
$7/2^{-}_{1}$&-40.07&-6.77&2.570&$0.863\pm0.087$\\
&&&2.40\cite{Hamamoto-2010}&0.224\cite{Hamamoto-2010}\\
$7/2^{-}_{2}$&-40.07&-6.77&4.592&$0.803\pm0.083$\\
&&&$\approx 4.592$\cite{Liu-2012}&$\approx 0.05$\cite{Liu-2012}\\
\bottomrule
\end{tabular}
\label{tbl3}
\end{table}


\section*{Declaration of competing interest}
  The authors state that they have no known competing financial or personal interests that could have influenced the work presented in this paper.
 
 \section*{Acknowledgements}
The authors gratefully acknowledge the computational facilities provided by Aliah University in Kolkata, India.


\begin{thebibliography}{0}
\bibitem{Panin-2021} V. Panin, T. Aumann, C. A. Bertulani, Quasi-free scattering in inverse kinematics as a tool to unveil the structure of nuclei, Eur. Phys. J. A 57 (2021) 103, 
\href{blue}{https://doi.org/10.1140/epja/s10050-021-00416-9}.
%
\bibitem{Kasuya-2021} H. Kasuya, K. Yoshida, Hartree–Fock–Bogoliubov theory for odd-mass nuclei with a time-odd constraint and application to deformed halo nuclei, Prog. Theor. Exp. Phys. 2021 (2021) 013D01,
\href{blue}{https://doi.org/10.1093/ptep/ptaa163}, arXiv: 2005.03276v2.

\bibitem{Shubhchintak-2014} Shubhchintak, R. Chatterjee, Deformation effects in the Coulomb breakup of $^{31}$Ne, Nucl. Phys. A 922 (2014) 99-111, 
\href{blue}{https://doi.org/10.1016/j.nuclphysa.2013.11.010}, arXiv: 1311.7040v1.

\bibitem{Liu-2002} Z. H. Liu, Halo Nature of $^{14,15}$C, Chin. Phys. Lett. 19 (2002) 1071-1073,
\href{blue}{https://doi.org/10.1088/0256-307X/19/8/313}.

\bibitem{Tanihata-1985} I. Tanihata, et al., Measurements of interaction cross sections and nuclear radii in the light p-shell region, Phys. Rev. Lett. 55 (1985) 2676, 
\href{blue}{https://doi.org/10.1103/PhysRevLett.55.2676}.

\bibitem{Fukuda-1991} M. Fukuda, et al., Neutron halo in $^{11}$Be studied via reaction cross sections,  Phys. Lett. B 268 (1991) 339,
\href{blue}{https://doi.org/10.1016/0370-2693(91)91587-L}.

\bibitem{Schmitt-2012} K. T. Schmitt, et. al., Halo Nucleus $^{11}$Be: A Spectroscopic Study via Neutron Transfer, Phys. Rev. Lett. 108 (2012) 192701,
\href{blue}{https://doi.org/10.1103/PhysRevLett.108.192701}.

\bibitem{Nakamura-1999} T. Nakamura, et al., Coulomb dissociation of $^{19}$C and its halo structure, Phys. Rev. Lett. 83 (1999) 1112, 
\href{blue}{https://doi.org/10.1103/PhysRevLett.83.1112}.

\bibitem{Yamaguchi-2011} T. Yamaguchi, et al., Nuclear reactions of $^{19,20}$C on a liquid hydrogen target measured with the superconducting TOF spectrometer, Nucl. Phys. A 864 (2011) 1-37,
\href{blue}{https://doi.org/10.1016/j.nuclphysa.2011.05.095}.

\bibitem{Khan-2021} M. A. Khan, et al., Application of supersymmetric quantum mechanics to calculate resonance energy and wave function of $^{19}$C halo nucleus, Few-Body Syst. 62 (2021) 54, 
\href{blue}{https://doi.org/10.1007/s00601-021-01640-1}. 

\bibitem{Nakamura-2014} T. Nakamura, et al., Deformation-driven p-wave halos at the drip line: $^{31}$Ne, Phys. Rev. Lett. 112 (2014) 142501, 
\href{blue}{https://doi.org/10.1103/PhysRevLett.112.142501}.

\bibitem{Shi-2014} M. Shi, et al., Relativistic extension of the complex scaling method for resonant states in deformed nuclei, Phys. Rev. C 90 (2014) 034319, 
\href{blue}{https://doi.org/10.1103/PhysRevC.90.034319}.

\bibitem{Watanabe-2014} S. Watanabe, et al., Ground-state properties of neutron-rich Mg isotopes, Phys. Rev. C 89 (2014) 044610,
\href{blue}{https://doi.org/10.1103/PhysRevC.89.044610}, arXiv: 1404.2373v1.

\bibitem{Tanihata-1992} I. Tanihata, et al., Determination of the density distribution and the correlation of halo neutrons in $^{11}$Li, Phys. Lett. B 287 (1992) 307, 
\href{blue}{https://doi.org/10.1016/0370-2693(92)90988-G}.

\bibitem{Dutta-2003} S. K. Dutta, T. K. Das, M. A. Khan, B. Chakrabarti, Computation of $2^{+}$ resonance in $^{6}$He: bound state in the continuum, J. Phys. G: Nucl. Part. Phys. 29 (2003) 2411,
\href{blue}{https://doi.org/10.1088/0954-3899/29/10/307}.

\bibitem{Souza-2016} L. A. Souza, et al., Core momentum distribution in two-neutron halo nuclei, Phys. Lett. B 757 (2016) 368-375, 
\href{blue}{https://doi.org/10.1016/j.physletb.2016.03.087}, arXiv: 1512.02070v1.

\bibitem{Ibraheem-2019} A. A. Ibraheem, et al., Elastic scattering of one-proton halo nucleus $^{17}$F on different mass targets using semi microscopic potentials, Rev. Mex. Fiz. 65 (2019) 168, 
\href{blue}{https://doi.org/10.31349/revmexfis.65.168}.

\bibitem{Togano-2016} Y. Togano, et al., Interaction cross section study of the two-neutron halo nucleus $^{22}$C, Phys. Lett. B 761 (2016) 412, 
\href{blue}{https://doi.org/10.1016/j.physletb.2016.08.062}. 

\bibitem{Hasan-2019} M. Hasan, et al., Construction of a one-parameter family of isospectral potential to study resonances in weakly bound halo nuclei, Jour. Phys. Conf. Ser. 1354 (2019) 012003, 
\href{blue}{https://doi.org/10.1088/1742-6596/1354/1/012003}.

\bibitem{Fukuda-1999} M. Fukuda, et al., Density distribution of $^{8}$B studied via reaction cross sections, Nucl. Phys. A 656 (1999) 209-228,
\href{blue}{https://doi.org/10.1016/S0375-9474(99)00308-5}. 

\bibitem{Ren-1996} Z. Ren, et al., One-proton halo in $^{26}$P and two-proton halo in $^{27}$S, Phys. Rev. C 53 (1996) R572(R), 
\href{blue}{https://doi.org/10.1103/PhysRevC.53.R572}.

\bibitem{Cai-2002} X. Z. Cai, et al., Existence of a proton halo in $^{23}$Al and its significance, Phys. Rev. C 65 (2002) 024610, 
\href{blue}{https://doi.org/10.1103/PhysRevC.65.024610}.

\bibitem{Tanaka-2005} K. Tanaka, et al., Nucleon density distribution of proton drip-line nucleus $^{17}$Ne, Eur. Phys. Jour. A 25 (2005) 221, 
\href{blue}{https://doi.org/10.1140/epjad/i2005-06-182-1}.

\bibitem{Audi-2003} G. Audi, et al., The Ame2003 atomic mass evaluation: (II). Tables, graphs and references, Nucl. Phys. A 729 (2003) 337-676, 
\href{blue}{https://doi.org/10.1016/j.nuclphysa.2003.11.003}.

\bibitem{Urata-2011} Y. Urata, et al., Ground state properties and Coulomb dissociation of the deformed halo nucleus $^{31}$Ne, Phys. Rev. C 83 (2011) 041303(R), 
\href{blue}{https://doi.org/10.1103/PhysRevC.83.041303}.

\bibitem{Takechi-2012} M. Takechi, et al., Interaction cross sections for Ne isotopes towards the island of inversion and halo structures of $^{29}$Ne and $^{31}$Ne, Phys. Lett. B 707 (2012) 357-361,
\href{blue}{https://doi.org/10.1016/j.physletb.2011.12.028}.

\bibitem{Nakamura-2009} T. Nakamura, et al., Halo structure of the island of inversion nucleus $^{31}$Ne, Phys. Rev. Lett. 103 (2009)  262501, 
\href{blue}{https://doi.org/10.1103/PhysRevLett.103.262501}.

\bibitem{Jurado-2007} B. Jurado, et al., Mass measurements of neutron-rich nuclei near the $N=20$ and $28$ shell closures, Phys. Lett. B 649 (2007) 43, 
\href{blue}{https://doi.org/10.1016/j.physletb.2007.04.006}.

\bibitem{Wapstra-2003} A. H. Wapstra, G. Audi, C. Thibault, The Ame2003 atomic mass evaluation: (I). Evaluation of input data, adjustment procedures, Nucl. Phys. A 729 (2003) 129-336,
\href{blue}{https://doi.org/10.1016/j.nuclphysa.2003.11.002}.

\bibitem{Gaudefroy-2012} L. Gaudefroy, et al., Direct mass measurements of $^{19}$B, $^{22}$C, $^{29}$F, $^{31}$Ne, $^{34}$Na and other light exotic nuclei, Phys. Rev. Lett. 109 (2012) 202503, 
\href{blue}{https://doi.org/10.1103/PhysRevLett.109.202503}.

\bibitem{Poves-1994} A. Poves, J. Retamosa, Theoretical study of the very neutron-rich nuclei around $N = 20$, Nucl. Phys. A 571 (1994) 221,
\href{blue}{https://doi.org/10.1016/0375-9474(94)90058-2}.

\bibitem{Descouvemont-1999} P. Descouvemont, Microscopic cluster study of the $^{31}$Ne and $^{32}$Ne nuclei, Nucl. Phys. A 655 (1999) 440-449,
\href{blue}{https://doi.org/10.1016/S0375-9474(99)00305-X}.

\bibitem{Ren-2001} Z. Ren, et al., Neutron halo and nuclear shell structure in new nuclide $^{31}$Ne, Commun. Theor. Phys. 35 (2001) 717, 
\href{blue}{https://doi.org/10.1088/0253-6102/35/6/717}.

\bibitem{Hamamoto-2010} I. Hamamoto, Interpretation of Coulomb breakup of $^{31}$Ne in terms of deformation, Phys. Rev. C 81 (2010) 021304(R), 
\href{blue}{https://doi.org/10.1103/PhysRevC.81.021304}, arXiv: 0912.4625v1.

\bibitem{Minomo-2012} K. Minomo, et al., Determination of the structure of $^{31}$Ne by a fully microscopic framework, Phys. Rev. Lett. 108 (2012) 052503,
\href{blue}{https://doi.org/10.1103/PhysRevLett.108.052503}. 

\bibitem{Liu-2012} Q. Liu, J. Y. Guo, Z. M. Niu, S. W. Chen, Resonant states of deformed nuclei in the complex scaling method, Phys. Rev. C 86 (2012) 054312, arXiv: 1211.6920v1.

\bibitem{Zhang-2014} S. S. Zhang, et al., Microscopic self-consistent study of neon halos with resonant contributions, Phys. Lett. B 730 (2014) 30, 
\href{blue}{https://doi.org/10.1016/j.physletb.2014.01.023}.

\bibitem{Tian-2017} Y. J. Tian, Q. Liu, T. H. Heng, J. Y. Guo, Research on the halo in $^{31}$Ne with the complex momentum representation method, Phys. Rev. C 95 (2017) 064329,
\href{blue}{https://doi.org/10.1103/PhysRevC.95.064329}, arXiv: 1707.00246v1. 

\bibitem{He-2019} F. He, Electric properties of the neon-31 nuclei under halo EFT formalism, Few-Body Syst. 60 (2019) 16,
\href{blue}{https://doi.org/10.1007/s00601-019-1484-1}.


\bibitem{Moiseyev-1998} Nimrod Moiseyev, Quantum theory of resonances: calculating energies, widths and cross-sections by complex scaling, Phys. Rep. 302 (1998) 212,
\href{blue}{https://doi.org/10.1016/S0370-1573(98)00002-7}

\bibitem{Descouvemont-1990} P. Descouvemont and M. Vincke, Iterative method for resonance properties in the R-matrix theory,
Phys. Rev. A 42, (1990) 3835, \href{blue}{https://doi.org/10.1103/PhysRevA.42.3835}


\bibitem{Kukulin-1989} V.I. Kukulin, V.M. Krasnopl’sky, J. Horácek, Theory of Resonances: Principles and Applications, Kluwer Academic, Dordrecht, 1989.
\href{blue}{https://www.worldcat.org/title/theory-of-resonances-principles-and-applications/oclc/858852621}

\bibitem{Baluja-1982}K.L.Baluja, P.G.Burke, L.A.Morgan, R-matrix propagation program for solving coupled second-order differential equations, Computer Physics Communications
Volume 27, Issue 3, September 1982, Pages 299-307, \href{blue}{https://doi.org/10.1016/0010-4655(82)90177-1}


\bibitem{Khan2-2021} M. A. Khan, et al., Hyperspherical three-body model calculation for the bound and resonant states of the neutron dripline nuclei $^{42,44}$Mg using isospectral potential, Nucl. Phys. A 1015 (2021) 122316, 
\href{blue}{https://doi.org/10.1016/j.nuclphysa.2021.122316}.

\bibitem{Johnson-1978} B. R. Johnson, The renormalized Numerov method applied to calculating bound states of the coupled‐channel Schroedinger equation, J. Chem. Phys. 69 (1978) 4678, 
\href{blue}{https://doi.org/10.1063/1.436421}.

\bibitem{Nieto-1984} M. M. Nieto, Relationship between supersymmetry and the inverse method in quantum mechanics, Phys. Lett. B 145 (1984) 208-210, 
\href{blue}{https://doi.org/10.1016/0370-2693(84)90339-3}.

\bibitem{Khare-1989} A. Khare, U. Sukhatme, Phase-equivalent potentials obtained from supersymmetry, J. Phys. A: Math. Gen. 22 (1989) 2847, 
\href{blue}{https://doi.org/10.1088/0305-4470/22/14/031}.

\bibitem{Cooper-1995} F. Cooper, A. Khare, U. Sukhatme, Supersymmetry and quantum mechanics, Phys. Rep. 251 (1995) 267, 
\href{blue}{https://doi.org/10.1016/0370-1573(94)00080-M}.

\bibitem{Pappademos-1993} J. Pappademos, U. Sukhatme, A. Pagnamenta, Bound states in the continuum from supersymmetric quantum mechanics, Phys. Rev. A 48 (1993) 3525, 
\href{blue}{https://doi.org/10.1103/PhysRevA.48.3525}, arXiv: hep-ph/9305336v1.

\bibitem{Darboux-1882} G. Darboux, On a proposition relative to linear equations, C. R. Acad. Sci. Paris 94 (1882) 1456-1459.
\href{blue}{https://arxiv.org/pdf/physics/9908003.pdf}

\bibitem{Pahlavani-2012} M. R. Pahlavani, S. A. Alavi, Solutions of Woods—Saxon potential with spin-orbit and centrifugal terms through Nikiforov—Uvarov method, Commun. Theor. Phys. 58 (2012) 739,
\href{blue}{https://doi.org/10.1088/0253-6102/58/5/19}.

\bibitem{Dutta-2018} S.K. Dutta, D. Gupta, S. K. Saha, Resonance state wave functions of $^{15}$Be using supersymmetric quantum mechanics, Phys. Lett. B 776 (2018) 464–467,
\href{blue}{https://doi.org/10.1016/j.physletb.2017.12.008}.

\end{thebibliography}

\end{document}